\documentclass{iopart}

\usepackage[english]{babel}

\usepackage{graphicx}

\begin{document}

\title{Positive Correlations in Tunneling through coupled Quantum Dots}

\author{G.~Kie{\ss}lich, H. Sprekeler, A.~Wacker, and E.~Sch{\"o}ll}
\address{Institut f{\"u}r Theoretische Physik, Technische Universit{\"a}t
  Berlin, D-10623 Berlin, Germany}

\ead{kieslich@physik.tu-berlin.de}

\begin{abstract}

Due to the Fermi-Dirac statistics of electrons the temporal correlations of tunneling events 
in a double barrier setup are typically negative. 
Here, we investigate the shot noise behavior of a system of two capacitively 
coupled quantum dot states by means of a Master equation model. In an asymmetric setup 
positive correlations in the
tunneling current can arise due to the bunching of tunneling events.
The underlying mechanism will be discussed in detail in terms of the current-current correlation 
function and the frequency-dependent Fano factor.

\end{abstract}


\section{Introduction}

Resonant tunneling of electrons through bound states in mesoscopic double barrier devices still
attracts wide research activities. Not only the current flow is of interest. Due to the granularity
of the electron charge the tunneling current is a discrete stochastic process which leads to 
shot noise in the current signal \cite{BLA00}. The spectral power density gives information
about temporal correlations which are not available by conductance measurements alone.
In the pioneering theoretical work of L. Y. Chen et al. \cite{CHE91} a suppression of the zero-frequency 
spectral power density $S$ with respect to the uncorrelated value
$S_P=2eI$ (Poissonian noise, $e>0$ is the elementary charge, $I$ is the average current) in tunneling
through double barriers  was obtained. This refers to negative correlations in the tunneling current 
due to Pauli's exclusion principle: if the bound state becomes occupied by tunneling of an electron, 
then a subsequent electron ``has to wait'' until the state becomes empty. Hence, negative
correlations are a fingerprint of the fermionic statistics of electrons. Contrarily, boson
statistics lead to positive correlations, e.g. in the intensity-intensity correlation 
function of emission of photons from
a black radiation source due to temporal bunching of emission events (\cite{BUE03} and references therein). 

For tunneling through quantum dots (QDs) Coulomb interaction generates an additional source of correlations. 
It was shown in Ref.~\cite{KIE03a,KIE03b} that for an asymmetric setup of the tunneling
rates in a system of two capacitively coupled QD states the
current-voltage characteristic exhibits negative differential conductance.
In this regime positive correlations in the tunneling current
are present (super-Poissonian shot noise).
Here, we give a detailed examination of this effect and show that its origin is related
to bunching of tunneling events due to Coulomb blocking.
Positive correlations in tunneling current were also investigated in a double barrier resonant
tunneling diode \cite{IAN98}, in coupled metallic QDs \cite{GAT02}, and in a 
single-electron transistor setup \cite{SAF03}.

\section{Model}

We consider two QD single-particle states with energies $E_{1/2}$ which are coupled 
to emitter/collector contacts
by tunneling rates $\Gamma_{E/C}^i$ ($i=$ 1,2), respectively. The emitter/collector contact is 
assumed to be in local equilibrium and the occupation is given by the Fermi functions $f_{E/C}(E)$ 
and chemical potentials 
$\mu_{E/C}$, respectively. An applied bias voltage $eV=\mu_E-\mu_C$ drives the QD system out of equilibrium.
The tunneling energy $E$ is given by the difference of energies before and after a tunneling process 
considering the charging energy $U$ which is present if both QD states are occupied. 
The time evolution of the occupation probabilities $P_\nu$ for the Fock states $\nu =(n_1,n_2)$ 
($n_i\in\{0,1\}$, $i=$ 1,2) is described by the Master equation 
$\dot{\underline{P}}=\mathbf{\underline{\underline{M}}}\,\underline{P}$. The 4$\times$4-matrix
$\mathbf{\underline{\underline{M}}}$ consists of the transition rates for tunneling into or out of the 
QD system which contain the occupation of the contacts and the tunneling rates 
(for details see \cite{KIE03b}).

The stationary current is given by 
$I=\sum_\nu\left(\mathbf{\underline{\underline{j}}_{E}}\,\underline{P}^0\right)_\nu$ with the stationary
occupation probability given by $\mathbf{\underline{\underline{M}}}\,\underline{P}^0=0$ and the
operator for the current at the emitter barrier $\mathbf{\underline{\underline{j}}_{E}}$, 
respectively. The current-current correlation function at the emitter barrier for $t\ge 0$ is 
$C_{EE}(t)\equiv\langle I_E(t)I_E(0)\rangle - I^2 =
\sum_\nu\left(\mathbf{\underline{\underline{j}}_{E}}\mathbf{\underline{\underline{T}}}(t)
\mathbf{\underline{\underline{j}}_{E}}\underline{P}^0\right)_\nu+\delta (t)eI - I^2$ 
with the time translation operator
$\mathbf{\underline{\underline{T}}}(t)\equiv\exp{\mathbf{\underline{\underline{M}}}t}$ so that 
$\underline{P}(t)=\mathbf{\underline{\underline{T}}}(t)\underline{P}(0)$ holds \cite{HER93}. 
Then, the spectral power density reads (Wiener-Khintchine theorem) $S_{EE}(\omega )= 2\int_0^{\infty}dt
C_{EE}(t)\cos{\omega t}$. As a measure for correlations at the emitter barrier the Fano factor is defined as 
$\alpha (\omega )\equiv\frac{S_{EE}(\omega )}{2eI}$ which is unity for uncorrelated tunneling, $< 1$ 
for negative correlations, and $> 1$ for positive correlations in the tunneling current (super-Poissonian 
shot noise).
Note that $S_{EE}(0)=S_{CC}(0)=S_{EC}(0)=S_{CE}(0)$, but  $S(\omega )\neq S_{EE}(\omega )$ for $\omega\neq 0$.

\section{Results and Discussion}

\begin{figure}[htb]
  \begin{center}
    \includegraphics[width=\textwidth]{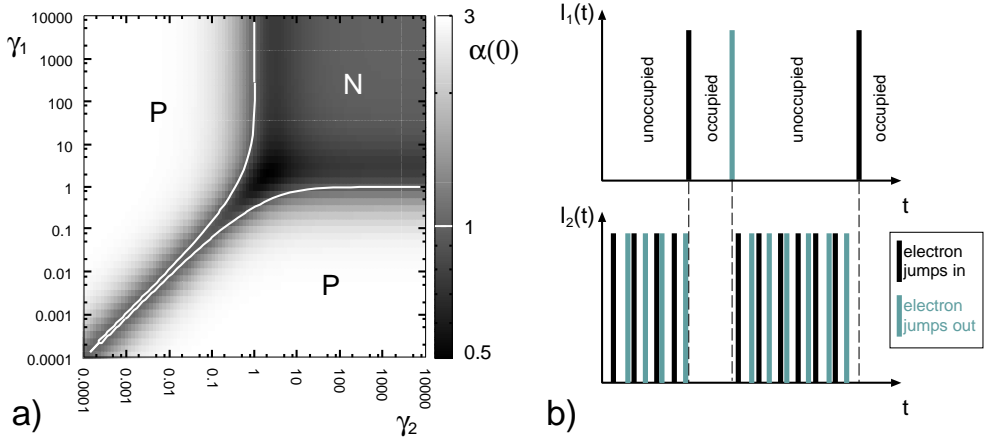}
    \caption{a) Fano factor $\alpha (0)$ vs. $\gamma_1=\Gamma_C^1/\Gamma_E$ and 
      $\gamma_2\equiv\Gamma_C^2/\Gamma_E$ 
      (white lines correspond to $\alpha (0)=1$ and 
      confine the regions $P$ and $N$). 
	b) Schematic realisations of current $I_i$ vs. time $t$ through 
	level $i=1$ (upper plot) and $i=2$ (lower plot) for $\gamma_1\ll\gamma_2=$1.}
  \end{center}
  \label{fig1}
\end{figure}

We consider the case where both energies of the QD single-particle states 
are equal: $E_1=E_2$, the tunneling rates to the emitter are equal: $\Gamma_E^1=\Gamma_E^2\equiv\Gamma_E$, 
and a bias voltage
$V$ is applied such that both single-particle states $(1,0)$ and $(0,1)$ can be occupied but the 
occupancy of the
doubly-occupied state $(1,1)$ is energetically forbidden. Then, the zero-frequency Fano factor is derived 
analytically to be

\begin{equation}
  \alpha (0)= 1+\frac{2\left[\gamma_1^2+\gamma_2^2-\gamma_1\gamma_2(2+\gamma_1+\gamma_2)\right]}
	 {\left(\gamma_1+\gamma_2+\gamma_1\gamma_2\right)^2}
	 \label{eq:fano}
\end{equation}

\noindent
with the ratios of collector and emitter tunneling rates 
$\gamma_i\equiv\Gamma_C^i/\Gamma_E$ ($i=$ 1,2). The dependence of the Fano factor on 
$\gamma_1$ and $\gamma_2$ is depicted in Fig.1a. 
The Fano factor is symmetric with respect
to an interchange of $\gamma_1$ and $\gamma_2$. Two different regions corresponding to $\alpha\le 1$  and 
$\alpha >1$ are labeled by $N$ and $P$, respectively.
The region $N$ indicates negative correlations in the tunneling current for $\gamma_i\ge 1$ 
($i=$ 1,2). In this regime, mutual Coulomb blocking of the single-particle states $(1,0)$ and 
$(0,1)$ is negligible since the levels are mostly empty in 
time average due to the stronger collector coupling. 
If either one ratio or both ratios become 
smaller than unity the tunneling is positively correlated (region $P$ in Fig. 1a) 
except for $\gamma_1\simeq\gamma_2$ where $\alpha$ approaches unity. Hence, the conditions for
positive correlations are: $\gamma_1<1$ or $\gamma_2<1$, and $\gamma_1\not=\gamma_2$.
To illustrate the effect of positive correlations we consider a realisation of time-dependent currents 
$I_1(t)$ and $I_2(t)$ through the levels 1 and 2, respectively,  in Fig. 1b. 
There, it is assumed that $\gamma_1\ll\gamma_2=$1. 
If an electron jumps from the emitter into level 1 (black peak) this level becomes occupied. 
Until the level becomes empty by tunneling of the electron out in the collector (grey peak) no electrons can
jump in level 2 (Coulomb blocking). 
Only if level 1 is empty, a current $I_2$ is flowing. Hence, a bunching of tunneling events
in current $I_2(t)$ occurs.

In Fig. 2 the frequency-dependent Fano factor $\alpha (\omega)$ and in the insets of Fig. 2 
the current-current correlation function $C_{EE}(t)$ at the emitter barrier are shown. The dashed curve
$\alpha (\omega)$ represents the case $\gamma_1=\gamma_2=$1. 
It exhibits a minimum at $\omega =$0 and approaches unity
for $\omega\rightarrow\pm\infty$. The corresponding correlation function in the left inset is negative for
all times, i.e. only negative correlations are present (region $N$ in Fig.1a). By lowering
$\gamma_1$ below one a maximum at $\omega =0$ and two minima symmetric to $\omega =$0 arise (dotted curve: 
$\gamma_1=$0.4, $\gamma_2=$1; full curve: $\gamma_1=$0.2, $\gamma_2=$1).
The correlation function (full curve in the right inset of Fig. 2) belonging to the full curve 
$\alpha (\omega )$ ($\alpha >$1, $P$-region in 
Fig. 1a)  has now two contributions with different signs: 
positive (dotted curve) and negative (dashed curve). The positive part of the correlation function 
is due to the bunching of tunneling events as discussed with respect to  Fig. 1b and the negative 
part is due to anti-bunching caused by Pauli's exclusion principle which is still present in the 
bunches of tunneling through the current 
carrying level. Therefore, the frequency-dependent Fano factor in Fig. 2 consists of a 
sum of two Lorentzians with either positive or negative sign and different FWHM corresponding to
the two time scales related to the respective collector tunneling rates.

\begin{figure}[htb]
  \label{fig2}
  \begin{center}
    \includegraphics[scale=.5]{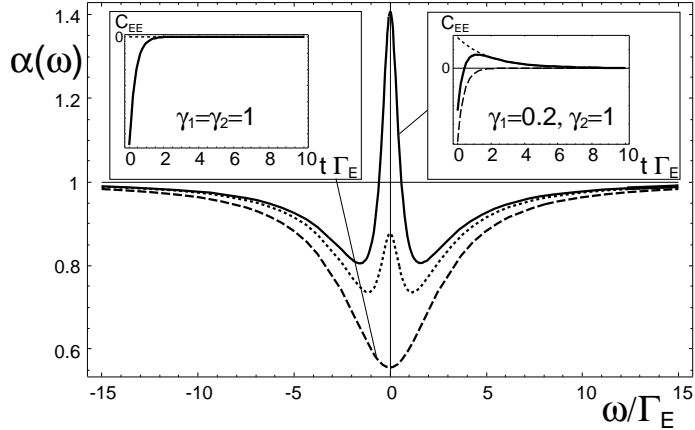}
    \caption{Fano factor $\alpha$ vs. frequency $\omega$ for $\gamma_1=1$ and $\gamma_2=1$ (dashed curve), 
      $\gamma_1=0.4$ (dotted curve), and $\gamma_1=0.2$ (full curve). Insets: 
      current-current correlation functions 
      at the emitter barrier $C_{EE}$ vs. time $t$ (full curves); dashed curve: negative part of $C_{EE}$, 
      dotted curve: 
      positive part of $C_{EE}$.}
  \end{center}
\end{figure}

\section{Conclusions}

We discussed the effect of positive correlations in the tunneling current through two single-particle states
in a regime where mutual Coulomb blocking  leads to bunching of tunneling events. We showed that the
current-current correlation function at one barrier consists of the sum of a 
negative and a positive term for an asymmetric
choice of the tunneling rates of both QD states. Then, the corresponding spectral power density is 
a sum of two Lorentzians with different FWHM and different signs which gives rise to a maximum in the 
zero-frequency Fano factor and which can become larger than one (super-Poissonian shot noise). 

\ack This work was supported by Deutsche Forschungsgemeinschaft in the framework of SFB 296.

\end{document}